\documentclass[aps,prl,twocolumn,superscriptaddress,footinbib,showpacs]{revtex4}
\usepackage{graphics,graphicx}
\usepackage{color}
\usepackage{amssymb,amsmath}
\usepackage{bookmath}
\usepackage{bm}

\begin{document}


\title{Pauli-limited upper critical field in dirty $d$-wave superconductors}
\author{A.~B.~Vorontsov}\altaffiliation{Present address: Dept. of Physics,
University of Wisconsin, Madison, WI}
\email[email:]{anton@physics.wisc.edu}
\author{I.~Vekhter}
\affiliation{Department of Physics and Astronomy,
             Louisiana State University, Baton Rouge, Louisiana, 70803, USA}
\author{M.~J.~Graf}
\affiliation{Theoretical Division, Los Alamos National Laboratory,
Los Alamos, NM 87545, USA.}
\date{\today}




\begin{abstract}
We calculate the Pauli-limited upper critical field and the
Fulde-Ferrell-Larkin-Ovchinnikov (FFLO) instability for {\it
dirty} $d$-wave superconductors within the quasiclassical theory
using the self-consistent $\hat{t}$-matrix approximation for
impurities. We find that the phase diagram depends sensitively on
the scattering rate and phase shift of nonmagnetic impurities. The
transition into the superconducting state is always second order
for weak (Born) scattering, while in the unitarity (strong)
scattering limit a first-order transition into both uniform and
spatially modulated superconducting states is stabilized. Contrary
to general belief, we find that the FFLO phase is robust against
disorder and survives impurity scattering equivalent to a $T_c$
suppression of roughly 40\%. Our results bear on the search
of FFLO states in heavy-fermion and layered organic
superconductors.
\end{abstract}

\maketitle




\paragraph*{Introduction.}
In type-II singlet superconductors a magnetic field suppresses
superconductivity for two reasons: (1) the phase of the Cooper
pair wave function couples to the vector potential resulting in
the appearance of vortices; (2) Zeeman coupling of the magnetic
field to the electron spins polarizes and splits the conduction
band, which destroys superconductivity when the loss in magnetic
energy equals the energy gain from pair
condensation\cite{cha62,clo62,sarma1963,maki1964}. This latter
mechanism is referred to as Pauli limiting and leads to a first or
second-order transition from the normal (N) to superconducting
(SC) state depending on the value of the magnetic field. It has
been predicted that a clean system at high fields can remain
superconducting beyond the Pauli limit by forming the nonuniform
Fulde-Ferrell-Larkin-Ovchinnikov (FFLO) state with a spatially
modulated order parameter \cite{ful64}. This state, however, is
suppressed by disorder \cite{asl68}.

In contrast to conventional (isotropic $s$-wave) superconductors,
unconventional ($d$-wave) superconductors are affected by
nonmagnetic impurities even at zero field; scattering averages the
gap over the Fermi surface and suppresses $T_c$. The different
rates of suppression of the uniform and FFLO states determine the
phase diagram in the field-temperature ($B$-$T$) plane. Agterberg
and Yang \cite{agt01} found that in two-dimensional (2D) $d$-wave
superconductors with purely Zeeman coupling, the N-SC transition
is of second order at all $T$, with the Larkin-Ovchinnikov (LO)
modulation, $\Delta_{LO} \sim \cos \vq\!\cdot\!\vR$, and the
uniform (USC) state, $\Delta_{USC}=const$, favored at low and high
temperatures, respectively, and a narrow intermediate $T$ region,
where the nodeless Fulde-Ferrell (FF) state, $\Delta_{FF} \sim
e^{i\vq\cdot\vR}$, is stabilized. Ref.~\cite{ada03} reported that
under combined orbital and Zeeman coupling in impure $d$-wave
superconductors the first-order transition into the vortex state
appears at intermediate temperatures. Very recently, Houzet and
Mineev \cite{hou06} studied orbital and impurity effects in
$s$-wave and $d$-wave Pauli-limited superconductors and concluded
that orbital effects are necessary for a first-order transition to
occur in 2D $d$-wave superconductors.
In contrast, for $s$-wave in 3D the
transition to the FFLO state is first-order\cite{buz97,3D}.

Remarkably, our understanding of impurity effects in nonuniform states is
still incomplete. In Refs.~\cite{agt01,ada03,hou06} the discussion
was limited to weak (Born) impurity scattering and focused only on
the Ginzburg-Landau (GL) regime close to the onset of the FFLO
instability, using an expansion in the modulation wave vector
$\vq$. However, $q=|\vq|$ increases rapidly to values comparable
to the inverse superconducting coherence length, $q\xi_0\sim 1$,
so this expansion quickly becomes invalid away from the critical
point.

In this Letter, we present a microscopic treatment of impurity
effects on the superconducting states in purely Pauli-limited
quasi-2D $d$-wave superconductors. Impurities are treated in the
self-consistent $\hat{t}$-matrix approximation (SCTA) covering the
weak (Born) and strong (unitarity) scattering limits
\cite{buchholtz1981}. The latter limit, never considered
previously, is especially important because of a
search for FFLO-like states in heavy-fermion and layered organic
superconductors \cite{bia03}, where impurity scattering is
strong~\cite{pet86}. Our approach is not limited to
an expansion in $q$, and hence is valid for any temperature and
impurity concentration along the second-order upper critical
field $B_{c2}$. We show that the phase diagram of a Pauli-limited
{\it dirty} $d$-wave superconductor is very different for
nonmagnetic impurities in the Born and unitarity limits. The
differences originate from the dependence on scattering strength
of quartic and higher order coefficients in the GL functional. The
first order N-SC transition, absent for Born scattering, is
stabilized by strong impurities, and is therefore expected in
heavy fermion systems.

\paragraph*{Quasiclassical equations.}
We follow Refs.~\cite{buchholtz1981,vorontsov2005} and solve the
quasiclassical equations for the $4\times4$-matrix Green's
functions in particle-hole and spin space, which satisfy the
normalization condition, $\whg^2 = -\pi^2\widehat{1}$, and the
transport equation,
    \bea \label{eq:QC} [ i\vare_m \widehat{\tau}_3
- \mu \vB \cdot \hat{\vS} - \whDelta(\vR, \hat{\vp}) -
\whs^{imp}(\vR; \vare_m) \,, \label{eq:eilZ} &&
 \\  \nonumber
\whg(\vR, \hat{\vp} ; \vare_m)] + i\hbar\vv_f(\hat{\vp}) \cdot
\gradR \; \whg(\vR, \hat{\vp}; \vare_m)  &=& 0 \,.
    \eea
Here $\mu$ is the magnetic moment, $\vare_m=\pi k_B T(2n+1)$ are
the Matsubara frequencies, $\whDelta$ is the mean-field
superconducting order parameter depending on the coordinate,
$\vR$, and momentum direction, $\hat{\vp}$, at the Fermi surface
with velocity $\vv_f$. The electron spin operator is $\hat{\vS} =
\vsigma \onehalf (1+\widehat\tau_3) + \vsigma^{*} \onehalf
(1-\widehat\tau_3)$. The Pauli matrices $\vsigma$ and $\bm{\tau}$
operate in spin and particle-hole space, respectively.
Eq.~(\ref{eq:eilZ}) is complemented by self-consistency equations
for $\whDelta$ and the impurity self-energy $\whs^{imp}$. We use
$\hbar = k_B = 1$.

In the SCTA  $\whs^{imp} = n_{imp} \widehat{t}$, with impurity
concentration  $n_{imp}$. For isotropic scattering the $t$-matrix
satisfies $ \widehat{t}(\vR; \vare_m) = u_0 \widehat{1} + u_0
\cN_f \langle \whg(\vR,\hvp;\vare) \rangle_{\hvp} \;
\widehat{t}(\vR; \vare_m)$, where angular brackets
$\langle\dots\rangle$ denote a normalized Fermi surface average.
The strength of the nonmagnetic impurity potential, $u_0$, is
expressed via the
isotropic scattering phase shift, $\delta_0 = \arctan (\pi
u_0 \cN_f)$; $\cN_f$ is the density of states per spin at the
Fermi surface. For Born (unitarity) scattering $\delta_0=0$
($\delta_0=\pi/2$) and the normal-state scattering rate $\Gamma
\equiv 1/2\tau_N=\Gamma_u \sin^2\delta_0$, with
$\Gamma_u=n_{imp}/\pi \cN_f$.

If we choose the direction of the spin quantization along $\vB=B
\hat{\vz}$ (which is allowed if the hamiltonian has spin-rotation
symmetry in the absence of the field), both $\whg$ and
$\whs^{imp}$ have block-diagonal structure corresponding to the
two spin projections. Hence, the quasiclassical equations for the
spin-up and spin-down sectors decouple~\cite{kle04}, and we solve
separately for the diagonal, $g_{s}$, and off-diagonal,
$f_{s},f'_{s} $, components of $\whg$, with $s = \pm1
\{\uparrow,\downarrow\}$, with the constraint $g_\sm{s}^2 -
f_\sm{s} f'_\sm{s} = -\pi^2$. However, both spin projections enter
the self-consistency equation for $\whDelta$. We assume a
separable pairing interaction $\cY(\hvp)\cY(\hvp^\prime)$, where
$\cY(\hvp)$ gives the angular dependence of the gap function with
the normalization $\langle \cY^2(\hvp)\rangle=1$.  For
$\Delta(\vR,\hvp) = \Delta(\vR) \cY(\hvp)$, we find

    \bea
\Delta(\vR) \ln {T\over T_{c0}} = T \sum_{\vare_m} \left(
 \left\langle \cF(\vR,\hvp;\vare_m) \right\rangle_{\hvp}
 - \frac{\pi\Delta(\vR)}{|\vare_m|}
\right) \,, \label{eq:sc}
\\
\whs^{imp}_\sm{s} = \cS_{s}
    \left( \begin{array}{cc}
    \cot\delta_0 + \langle g_\sm{s}\rangle/\pi & \langle f_\sm{s}\rangle/\pi \\
     \langle f'_\sm{s} \rangle/\pi & \cot\delta_0 - \langle g_\sm{s}\rangle/\pi
    \end{array} \right) \,.
    \label{eq:explicit_imp}
    \eea
Here $\cF(\vR,\hvp;\vare_m) = \onehalf \cY(\hvp)
 [{f_{\uparrow}(\vR, \hvp;\vare_m)+f_{\downarrow}(\vR, \hvp; \vare_m)}]$
and $\cS_{s}= {\Gamma}/[{1-\pi^{-2}{\sin^2\delta_0}
    (\langle g_\sm{s}\rangle^2-\langle f_\sm{s}\rangle \langle {f'_\sm{s}} \rangle + \pi^2 )}]$.
To calculate the $B$-$T$ phase diagram, we derive the
Ginzburg-Landau functional (expansion in $\Delta$ for arbitrary
$q$) by taking $\Delta(\vR) = \sum_\vq \Delta_\vq
\exp(i\vq\!\cdot\!\vR)$ and solving
Eqs.~(\ref{eq:QC})-(\ref{eq:explicit_imp}) together with the
normalization condition for $\whg$ to third order in
$\Delta$. We substitute the $n$-th order solutions
$f^{(1)}_\sm{s}, f^{(3)}_\sm{s}$ into Eq.~(\ref{eq:sc}) to obtain
the GL free energy difference between the SC and N states,
    \begin{widetext}
\begin{subequations}
\bea &&\Del\Omega^{GL} = \sum_\vq \alpha(T,B;\vq) |\Delta_\vq|^2 +
\sum_{\vq_1 \vq_2 \vq_3 \vq_4} \onehalf\beta(T,B; \vq_1, \vq_2;
\vq_3, \vq_4)  \; \Delta_{\vq_1} \Delta_{\vq_2} \Delta_{\vq_3}^*
\Delta_{\vq_4}^* \delta_{\vq_1+\vq_2,\vq_3+\vq_4}
\,, 
\\
 && \alpha(T,B;\vq) = \ln {T \over T_{c0}} -
2\pi T \sum_{\vare_m > 0} {\rm Re}
 \left(
   \left< \cY {\tilde\cY_\vq}{D_\vq^{-1}}\right> - {\vare_m^{-1}}
 \right)
\,, \label{eq:GL2}
\\
&& \beta(T,B; \vq_1, \vq_2; \vq_3, \vq_4) = \pi T \sum_{\vare_m >
0} {\rm Re}\left\{ \left< \tilde\cY_{\vq_1} \tilde\cY_{\vq_2}
\tilde\cY_{\vq_3} \tilde\cY_{\vq_4}
\frac{D_{(\vq_1+\vq_2+\vq_3+\vq_4)/4}} {D_{\vq_1} D_{\vq_2}
D_{\vq_3} D_{\vq_4}} \right> - \Gamma \Upsilon_{\vq_1 \vq_2 \vq_3
\vq_4} \right\}\,, \label{eq:GL4}
\\
\nonumber && \Upsilon_{\vq_1 \vq_2 \vq_3 \vq_4}= \left(\onehalf -
\sin^2\delta_0\right) \left( \Theta^{(2)}_{\vq_1
\vq_3}\Theta^{(2)}_{\vq_2 \vq_4}+\Theta^{(2)}_{\vq_1
\vq_4}\Theta^{(2)}_{\vq_2 \vq_3} \right)- \sin^2\delta_0 \left(
    \Theta^{(1)}_{\vq_1}\Theta^{(1)}_{\vq_3}\Theta^{(2)}_{\vq_2 \vq_4}\right.
\\
&&\left. \qquad\qquad\qquad\qquad
    +
    \Theta^{(1)}_{\vq_1}\Theta^{(1)}_{\vq_4}\Theta^{(2)}_{\vq_2 \vq_3}+
    \Theta^{(1)}_{\vq_2}\Theta^{(1)}_{\vq_3}\Theta^{(2)}_{\vq_1 \vq_4}+
    \Theta^{(1)}_{\vq_2}\Theta^{(1)}_{\vq_4}\Theta^{(2)}_{\vq_1 \vq_3}-
    2\Theta^{(1)}_{\vq_1}\Theta^{(1)}_{\vq_2}\Theta^{(1)}_{\vq_3}\Theta^{(1)}_{\vq_4}
    \right)\,,
    \label{eq:GL5}
    \eea
    \label{eq:GL}
\end{subequations}
\begin{figure*}[t]
\begin{center}
\begin{minipage}[h]{0.34\linewidth}
\caption{(Color online) The critical fields of 
Pauli-limited {\it dirty} $d$-wave superconductors
for transitions to USC (first-order, dashed) and FFLO states
(second-order, solid) for $\vq$ along nodes (squares) and
antinodes (circles). Superconductivity sets in at the highest
$B_{c2}$. (a) Pure case: Below $T_P=T_{\vq\ne0}$ the N-USC
transition is below the second-order N-FFLO transition. (b-d) Born
and unitarity impurities split $T_P$ and $T_{\vq\ne0}$, thus
$T_P^B < T_{\vq\ne0} < T_P^u$. A modulation along antinodes
quickly suppresses $B_{c2}$ relative to $\vq\| node$, panel (b).
For Born impurities the second-order FFLO transition is above the
N-USC line at all temperatures, as for the clean case. In the
unitarity limit the first-order N-USC transition preempts a N-LO
transition, see panel (d).} \label{fig:dwave}
\end{minipage}\hfill
\begin{minipage}[h]{0.64\linewidth}
\centerline{\includegraphics[width=0.85\linewidth]{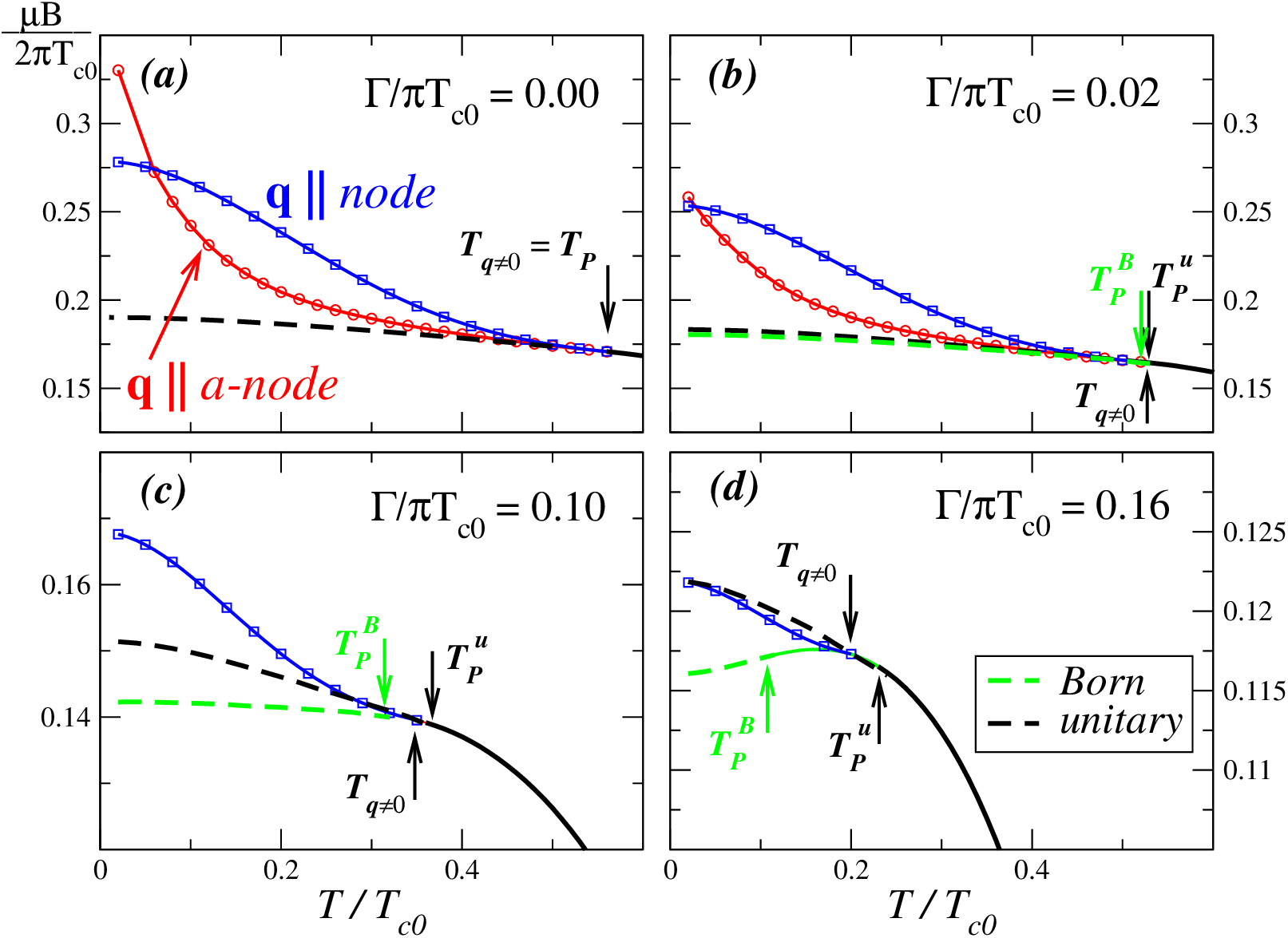}}
\end{minipage}
\end{center}
\end{figure*}
\end{widetext}
where we introduced the angular averages
$\Theta^{(1)}_{\vq_i}=\left<\tilde\cY_{\vq_i}D_{\vq_i}^{-1}\right>$,
$\Theta^{(2)}_{\vq_i
\vq_j}=\left<\tilde\cY_{\vq_i}\tilde\cY_{\vq_j}D_{\vq_i}^{-1}D_{\vq_j}^{-1}
 \right>$,
and defined $\eta_\vq = \onehalf \vv_f \cdot \vq$, and $D_\vq =
\vare_m + \Gamma + i(\mu B +\eta_\vq)$. We introduced
$\tilde\cY_\vq = \cY + \cY_{i,\vq}$ with 
$\cY_{i,\vq} = \Gamma \Theta^{(1)}_{\vq}$.

\paragraph*{Results.}
The second-order N-SC transition is determined from the GL
coefficient $\alpha(T,B;\vq) = 0$ and depends only on $\Gamma$,
but not on $\delta_0$. Thus the transition line is independent of
the phase shift. An instability into the modulated FFLO state
becomes possible below $T_{\vq\ne0}$, where the maximal $B_{c2}$
is found for $q\ne0$. This occurs when the GL coefficient $\kappa$
in the $q$-expansion of  $\alpha(T,B;\vq) \approx \alpha_0 +
\kappa q^2$ becomes negative,
    \be \kappa = 2\pi T \; {\rm Re}\,
\sum_{\vare_m>0} \frac{ 1 }{D_{\bf 0}^3} \left( \langle \cY^2
\eta_{\vq}^2 \rangle + \frac{ \Gamma \langle \cY \eta_{\vq}
\rangle^2 }{\vare_m+i\mu B } \right) \,. \label{eq:cK}
    \ee
For $d$-wave SC the last term 
vanishes, since $\langle \cY \eta_\vq \rangle = 0$. \indent
 In
contrast, the quartic term in the GL functional explicitly depends
on the scattering phase shift, Eqs.~(\ref{eq:GL4})-(\ref{eq:GL5}).
For example, it controls the location of the first-order
transition to the USC state, $T_P$, which competes with the FFLO
instability. In unconventional superconductors
$\langle\cY\rangle=\cY_{i,{\bf 0}} = 0$ and the critical point
$T_P$ is determined by a sign change of the GL coefficient $\beta$
at $\vq=0$,
     \be
\beta_0 = \pi T \; {\rm Re} \sum_{\vare_m>0} \left( \frac{ \langle
\cY^4 \rangle }{D_{\bf 0}^{3}}
 - \frac{ \Gamma (1-2\sin^2\delta_0) }{D_{\bf 0}^{4}}
\right)
 \,.
\label{eq:beta}
     \ee
For $\Gamma=0$, both $\kappa$ and $\beta_0$ become negative at
exactly the same temperature, $T_{\vq\ne0} = T_P\simeq 0.5615
T_{c0}$. Since the transition into the FFLO state has a higher
critical field at any temperature $T<T_P$, the first-order
transition is superseded by the onset of the FFLO
state~\cite{buz97}.

A comparison of $\kappa$ and $\beta_0$ shows that in {\it dirty}
unconventional superconductors $T_{\vq\ne0} = T_P$ only for
$\delta_0=\pi/4$. For Born (B) and unitarity (u) scattering
$\beta_0$ depends on $\delta_0$, such that $T_P^B$ and $T_P^u$
shift in opposite directions relative to $T_{\vq\ne0}$, hence
$T_P^B < T_{\vq\ne0} < T_P^u$ as shown in Fig.\ref{fig:dwave}. The
latter inequality is especially important since it shows that for
strong scatterers Pauli limiting leads to a first-order transition
into the USC state at high fields/low temperatures in the $B$-$T$
phase diagram. As the system becomes dirtier, i.e., the lifetime
$\tau_N$ decreases, these characteristic temperatures are
suppressed to zero in the following order, $T_P^B \to 0$ at
$\Gamma/\pi T_{c0} \gtrsim 0.18$, $T_{\vq\ne0}\to0$ at $\Gamma/\pi
T_{c0} \gtrsim 0.20$, and $T_P^u \to 0$ for $\Gamma/\pi T_{c0}
\gtrsim 0.22$. Note that for larger $\Gamma$ the N-USC transition
line is of second order at all $T$.

Fig.~\ref{fig:dwave} gives the upper critical field lines for
different states. Second-order transition lines are found by the
largest spatial modulation vector $q \equiv Q$ that maximize
{$B_{c2}$.} In clean $d$-wave
SC~\cite{mak96,shi97,yan98,vorontsov2005} the modulation is along
a gap maximum (antinode) at low $T/T_{c0}<0.06$, and along a gap
node for $0.06 < T/T_{c0}<0.56$, see Fig. \ref{fig:dwave}(a).
However, already for small impurity scattering, $\Gamma/\pi T_{c0}
\gtrsim 0.02$, the critical field for $\vq || antinode$ is lowered
below $B_{c2}^{\vq || node}$, and the stable configuration is with
$\vq\| node$ over the entire range of existence of the FFLO state,
see Fig. \ref{fig:dwave}(b).

Determining the first-order transition lines of $B_{c2}$ requires
a self-consistent calculation of the full free energy functional,
the details of which will be given elsewhere~\cite{abv08}. We find
that in the Born limit the first-order transition is always below
$B_{c2}^{FFLO}$, in agreement with \cite{agt01,hou06}. In
contrast, in the unitarity limit $T_{\vq\ne0} < T_P^u$ and
$B_{c2}^{FFLO}$ is below the first-order transition to the USC
state, see Figs.~\ref{fig:dwave}(b-d).

\begin{figure}[t]
\centerline{\includegraphics[width=0.90\linewidth]{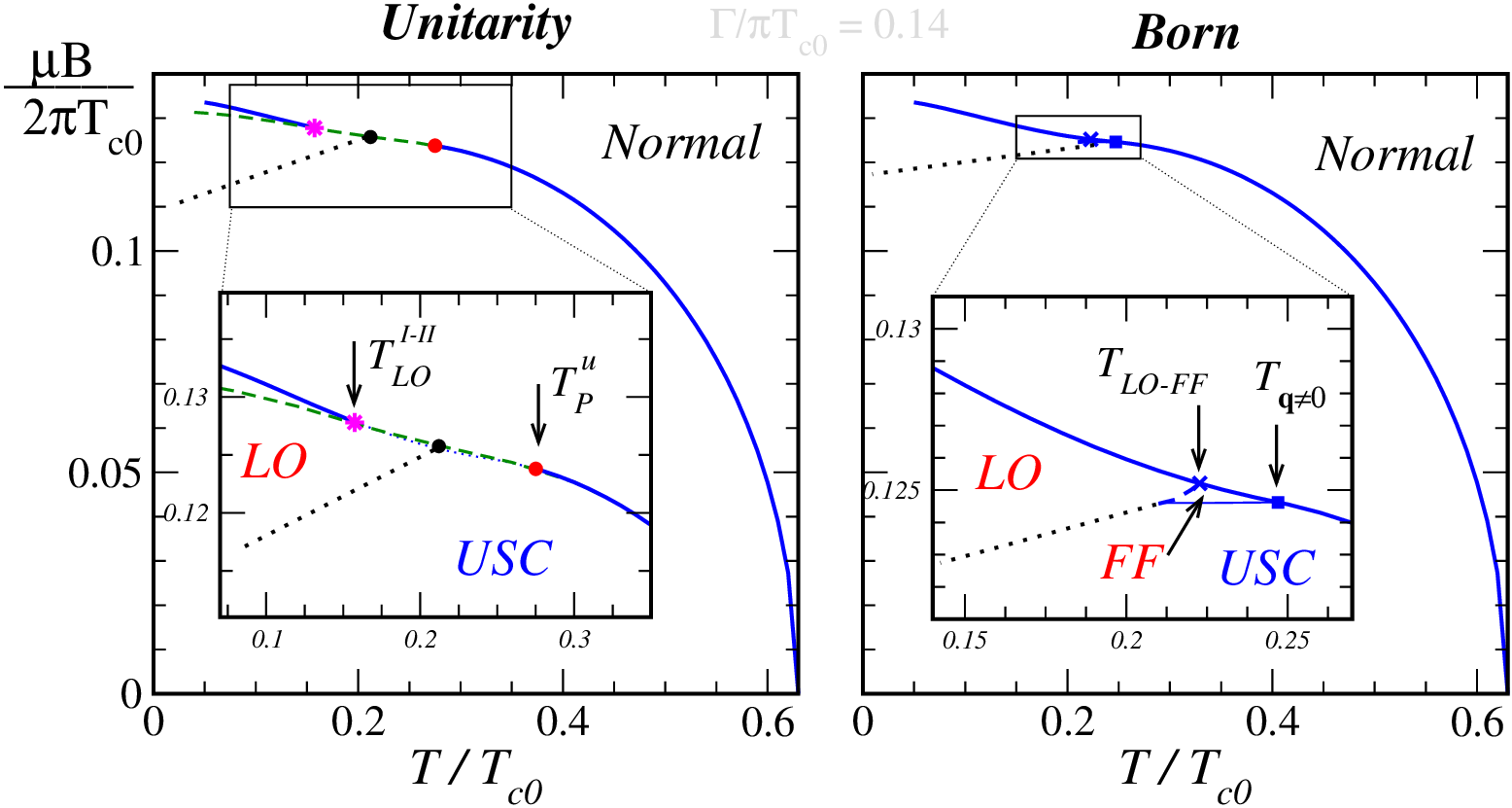}}
\caption{(Color online) The phase diagram
for $\Gamma / \pi T_{c0} = 0.14$. Left panel: transition into the
LO state at low $T$ becomes first order above $T_{LO}^{I-II}$ for
unitarity impurities. There is also a region of a first-order
transition into the USC state below $T_P$. Right panel: Born
impurities result in second-order transitions. The LO state is
favored in large parts of the phase diagram over the FF state,
except near $T_{\vq\ne0}$~\cite{agt01,hou06}. The transition line
shown between the LO and USC states (dotted line) is qualitative.
} \label{fig:PD}
\end{figure}

For intermediate impurity scattering, the phase diagram is given
in Fig.~\ref{fig:PD}. To determine the structure of the SC state
near $B_{c2}$, we analyze the GL free energy, Eq.(\ref{eq:GL}),
for four possible phases: USC [$\Delta(\vR)=\Delta_{USC}$],
FF with a single Fourier component $\vQ_1 = (Q,0)$
[$\Delta(\vR)=\Delta_{FF}\exp(iQx)$], LO with $\{\vQ_1,\vQ_3\} =
\{(\pm Q,0)\}$ [$\Delta(\vR)=\Delta_{LO}2\cos Qx$], and square
lattice (SQ) with $\{\vQ_1,\vQ_3, \vQ_2,\vQ_4 \} = \{(\pm Q,0),
(0,\pm Q)\}$ [$\Delta(\vR)=\Delta_{SQ}2(\cos Qx + \cos Qy)$]. The
$x,y-$axes are along the gap nodes. For each phase, we calculate
$\Del\Omega^{GL}_i = - \alpha^2/\beta_i$, with $\beta_{FF} = 2
\beta_{1111}$, $\beta_{LO} = \beta_{1111} + 2 \beta_{1313}$ and
$\beta_{SQ} = 0.5 (\beta_{1111} + 2 \beta_{1212} + 2 \beta_{1313}
+ 2 \beta_{1414} + 2 \beta_{1324} )$, where $\beta_{ijkl} =
\beta(T,B; \vQ_i, \vQ_j; \vQ_k, \vQ_l)$. Along the second-order
transition line the phase with the lowest positive value of
$\beta$ has the lowest energy.


For Born impurities (Fig. \ref{fig:PD} right), $\beta_i > 0$
for all nonuniform states, and the LO state is favored in most of
the phase diagram except a small region below $T_{\vq\ne0}$, where
the FF phase is stabilized for the impure case
~\cite{agt01,hou06}. Analysis of $\Delta\Omega^{GL}$ indicates
that this phase is separated by a second-order transition from the
USC and by first-order from the LO state.

The situation is very different for strong impurities
(Fig.~\ref{fig:PD} left). Following the $B_{c2}$ line from $T_c
(B=0)$ to lower $T$, we reach the critical point $T_P$, below
which the N-USC transition is of first order. At $T\to0$ the
transition is second order into the LO state, but becomes first
order above $T_{LO}^{I-II}$. We estimate where the first-order
N-USC and N-LO lines meet. However, determining the location of
this point and the LO-USC transition line requires a fully
self-consistent treatment of the nonuniform problem
\cite{vorontsov2005,bur94}, which is beyond the scope of this
work.


\begin{figure}[t]
\centerline{\includegraphics[width=0.90\linewidth]{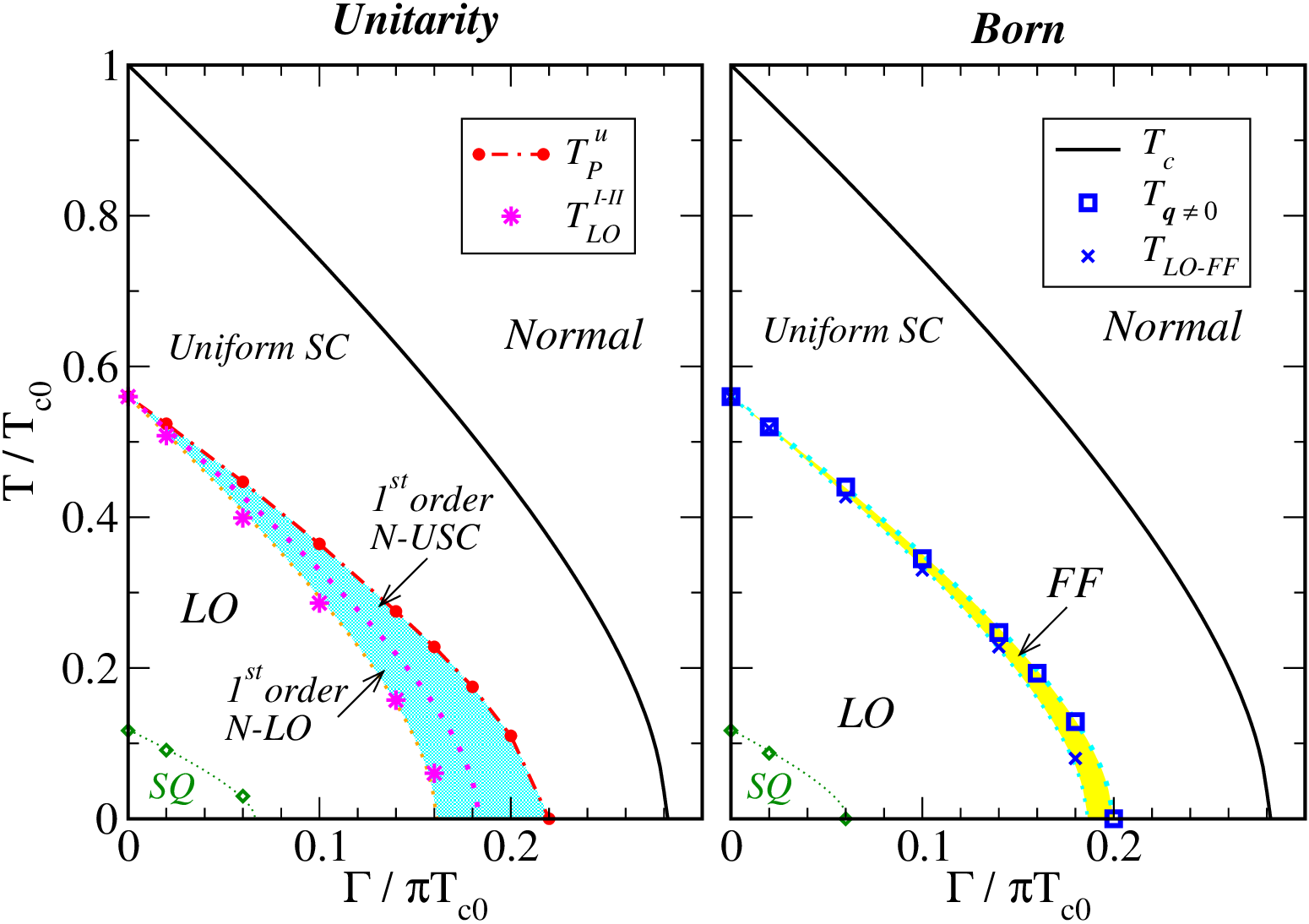}}
\caption{(Color online) Phase transitions along the $B_{c2}$
line in the $\Gamma$-$T_c$ plane. The (black) solid line is
$T_c(\Gamma)$ at $B=0$. For Born impurities the LO state exists
for $T<T_{\vq\ne0}$, with a small region occupied by the FF state.
For unitarity scattering a first-order transition appears into the
USC and LO states. At low $T$ a square lattice FFLO state (SQ)
\cite{shi97} is rapidly suppressed with increasing $\Gamma$
\cite{wan07}. } \label{fig:concl}
\end{figure}


\paragraph*{Conclusions}
We summarize our results in Fig.~\ref{fig:concl}, where we show
all states that arise along the upper critical field line for
fixed $\Gamma$. For nonuniform states, we only consider
modulations along gap nodes, since states with $\vq || antinode$
are destabilized even faster by impurities. We find for {\it
dirty} $d$-wave superconductors that the FFLO state is quite
robust and survives impurity scattering equivalent 
to $\sim 40$\% of $T_c$ suppression or a mean-free path $\ell$ of
$\xi_0/\ell \alt 0.16$.
This result is important for the search of
an FFLO state in doped Ce-115 \cite{doped115}, and other
heavy-fermion and layered organic superconductors.

Notably, the differences between weak and strong impurity
scattering are significant. In the Born limit $T_P$ is suppressed
below the onset of the nonuniform state, $T_P^B < T_{\vq\ne0}$,
and the transition is always of second order. Impurities stabilize
a narrow region of the Fulde-Ferrell state just below
$T_{\vq\ne0}$. In contrast in the unitarity limit
(relevant to recent experiments) $T_{\vq\ne0} < T_P^u$ and the
first-order transition into the uniform state preempts a modulated
state. Below $T\sim T_{\vq\ne0}$ the transition into the
Larkin-Ovchinnikov state begins as a first-order line and becomes
second-order at lower $T$. 
Importantly, in this limit the interplay of Zeeman splitting and disorder,
even without orbital effects,
drives the transition between the normal and superconducting
state first order.

We acknowledge support from the Louisiana Board of Regents
(A.B.V.\ and I.V.) and the US Dept.\ of Energy at LANL, contract
no.\ DE-AC52-06NA25396, (M.J.G.).




\begin{thebibliography}{25}
\expandafter\ifx\csname natexlab\endcsname\relax\def\natexlab#1{#1}\fi
\expandafter\ifx\csname bibnamefont\endcsname\relax
  \def\bibnamefont#1{#1}\fi
\expandafter\ifx\csname bibfnamefont\endcsname\relax
  \def\bibfnamefont#1{#1}\fi
\expandafter\ifx\csname citenamefont\endcsname\relax
  \def\citenamefont#1{#1}\fi
\expandafter\ifx\csname url\endcsname\relax
  \def\url#1{\texttt{#1}}\fi
\expandafter\ifx\csname urlprefix\endcsname\relax\def\urlprefix{URL }\fi
\providecommand{\bibinfo}[2]{#2}
\providecommand{\eprint}[2][]{\url{#2}}



\bibitem[{\citenamefont{Chandrasekhar}(1962)}]{cha62}
\bibinfo{author}{\bibfnamefont{B.}~\bibnamefont{Chandrasekhar}},
  \bibinfo{journal}{Appl. Phys. Lett.} \textbf{\bibinfo{volume}{1}},
  \bibinfo{pages}{7} (\bibinfo{year}{1962}).


\bibitem[{\citenamefont{Clogston}(1962)}]{clo62}
\bibinfo{author}{\bibfnamefont{A.}~\bibnamefont{Clogston}},
  \bibinfo{journal}{Phys. Rev. Lett.} \textbf{\bibinfo{volume}{9}},
  \bibinfo{pages}{266} (\bibinfo{year}{1962}).



\bibitem[{\citenamefont{Sarma}(1963)}]{sarma1963}
\bibinfo{author}{\bibfnamefont{G.}~\bibnamefont{Sarma}},
  \bibinfo{journal}{J. Phys. Chem. Solids} \textbf{\bibinfo{volume}{24}},
  \bibinfo{pages}{1029} (\bibinfo{year}{1963}).



\bibitem[{\citenamefont{Maki and Tsuneto}(1964)}]{maki1964}
\bibinfo{author}{\bibfnamefont{K.}~\bibnamefont{Maki}} \bibnamefont{and}
  \bibinfo{author}{\bibfnamefont{T.}~\bibnamefont{Tsuneto}},
  \bibinfo{journal}{Prog. Theor. Phys.} \textbf{\bibinfo{volume}{31}},
  \bibinfo{pages}{945} (\bibinfo{year}{1964}).



\bibitem[{\citenamefont{Fulde and Ferrell}(1964)}]{ful64}
\bibinfo{author}{\bibfnamefont{P.}~\bibnamefont{Fulde}} \bibnamefont{and}
  \bibinfo{author}{\bibfnamefont{R.}~\bibnamefont{Ferrell}},
  \bibinfo{journal}{Phys. Rev.} \textbf{\bibinfo{volume}{135}},
  \bibinfo{pages}{A550} (\bibinfo{year}{1964});
\bibinfo{author}{\bibfnamefont{A.~I.} \bibnamefont{Larkin}} \bibnamefont{and}
  \bibinfo{author}{\bibfnamefont{Y.~N.} \bibnamefont{Ovchinnikov}},
  \bibinfo{journal}{Zh. Eskp. Teor. Fiz.} \textbf{\bibinfo{volume}{47}},
  \bibinfo{pages}{1136} (\bibinfo{year}{1964}), \bibinfo{note}{[Sov. Phys. JETP
  {\bf 20}, 762 (1965)]}.



\bibitem[{\citenamefont{Aslamazov}(1968)}]{asl68}
\bibinfo{author}{\bibfnamefont{L.~G.}~\bibnamefont{Aslamazov}},
  \bibinfo{journal}{Sov. Phys. JETP} \textbf{\bibinfo{volume}{28}},
  \bibinfo{pages}{773} (\bibinfo{year}{1969}).





\bibitem[{\citenamefont{Agterberg and Yang}(2001)}]{agt01}
\bibinfo{author}{\bibfnamefont{D.}~\bibnamefont{Agterberg}} \bibnamefont{and}
  \bibinfo{author}{\bibfnamefont{K.}~\bibnamefont{Yang}}, \bibinfo{journal}{J.
  Phys.: Cond. Matt.} \textbf{\bibinfo{volume}{13}}, \bibinfo{pages}{9259}
  (\bibinfo{year}{2001}).



\bibitem[{\citenamefont{Adachi and Ikeda}(2003)}]{ada03}
\bibinfo{author}{\bibfnamefont{H.}~\bibnamefont{Adachi}} \bibnamefont{and}
  \bibinfo{author}{\bibfnamefont{R.}~\bibnamefont{Ikeda}},
  \bibinfo{journal}{Phys. Rev. B} \textbf{\bibinfo{volume}{68}},
  \bibinfo{pages}{184510} (\bibinfo{year}{2003}).



\bibitem[{\citenamefont{Houzet and Mineev}(2006)}]{hou06}
\bibinfo{author}{\bibfnamefont{M.}~\bibnamefont{Houzet}} \bibnamefont{and}
  \bibinfo{author}{\bibfnamefont{V.~P.} \bibnamefont{Mineev}},
  \bibinfo{journal}{Phys. Rev. B} \textbf{\bibinfo{volume}{74}},
  \bibinfo{pages}{144522} (\bibinfo{year}{2006}).



\bibitem[{\citenamefont{Buzdin and Kachkachi}(1997)}]{buz97}
\bibinfo{author}{\bibfnamefont{A.}~\bibnamefont{Buzdin}} \bibnamefont{and}
  \bibinfo{author}{\bibfnamefont{H.}~\bibnamefont{Kachkachi}},
  \bibinfo{journal}{Phys. Lett. A} \textbf{\bibinfo{volume}{225}},
  \bibinfo{pages}{341} (\bibinfo{year}{1997}).

\bibitem{3D}
\bibinfo{author}{\bibfnamefont{S.}~\bibnamefont{Matsuo}, et.al.,}
  \bibinfo{journal}{J. Phys. Soc. Japan} \textbf{\bibinfo{volume}{67}},
  \bibinfo{pages}{280} (\bibinfo{year}{1998});
\bibinfo{author}{\bibfnamefont{C.}~\bibnamefont{Mora}} \bibnamefont{and}
\bibinfo{author}{\bibfnamefont{R.}~\bibnamefont{Combescot}},
  \bibinfo{journal}{Phys. Rev. B} \textbf{\bibinfo{volume}{71}},
  \bibinfo{pages}{214504} (\bibinfo{year}{2005}).



\bibitem[{\citenamefont{Buchholtz and Zwicknagl}(1981)}]{buchholtz1981}
\bibinfo{author}{\bibfnamefont{L.~J.} \bibnamefont{Buchholtz}} \bibnamefont{and}
  \bibinfo{author}{\bibfnamefont{G.}~\bibnamefont{Zwicknagl}},
  \bibinfo{journal}{Phys. Rev. B} \textbf{\bibinfo{volume}{23}},
  \bibinfo{pages}{5788} (\bibinfo{year}{1981});
  \bibinfo{author}{\bibfnamefont{S.}~\bibnamefont{Schmitt-Rink}},
  \bibinfo{author}{\bibfnamefont{K.}~\bibnamefont{Miyake}}, \bibnamefont{and}
  \bibinfo{author}{\bibfnamefont{C.}~\bibnamefont{Varma}},
  \bibinfo{journal}{Phys. Rev. Lett.} \textbf{\bibinfo{volume}{57}},
  \bibinfo{pages}{2575} (\bibinfo{year}{1986}); \bibinfo{author}{\bibfnamefont{P.~J.} \bibnamefont{Hirschfeld}},
  \bibinfo{author}{\bibfnamefont{D.}~\bibnamefont{Vollhardt}},
  \bibnamefont{and} \bibinfo{author}{\bibfnamefont{P.}~\bibnamefont{W\"olfle}},
  \bibinfo{journal}{Sol. State Comm.} \textbf{\bibinfo{volume}{59}},
  \bibinfo{pages}{111} (\bibinfo{year}{1986}).



\bibitem[{\citenamefont{Bianchi et al.}(2003)}]{bia03}
\bibinfo{author}{\bibfnamefont{A.}~\bibnamefont{Bianchi}} \bibnamefont{et al.},
  \bibinfo{journal}{Phys. Rev. Lett.} \textbf{\bibinfo{volume}{91}},
  \bibinfo{pages}{187004} (\bibinfo{year}{2003});
  \bibinfo{author}{\bibfnamefont{C.}~\bibnamefont{Capan}} \bibnamefont{et al.},
  \bibinfo{journal}{Phys. Rev. B} \textbf{\bibinfo{volume}{70}},
  \bibinfo{pages}{134513} (\bibinfo{year}{2004});
\bibinfo{author}{\bibfnamefont{J.}~\bibnamefont{Singleton}} \bibnamefont{et al.},
  \bibinfo{journal}{J. Phys.: Condens. Matter} \textbf{\bibinfo{volume}{12}},
  \bibinfo{pages}{L641} (\bibinfo{year}{2000});
\bibinfo{author}{\bibfnamefont{Y.}~\bibnamefont{Matsuda}} \bibnamefont{and}
  \bibinfo{author}{\bibfnamefont{H.}~\bibnamefont{Shimahara}},
  \bibinfo{journal}{J. Phys. Soc. Japan} \textbf{\bibinfo{volume}{76}},
  \bibinfo{pages}{051005} (\bibinfo{year}{2007});
\bibinfo{author}{\bibfnamefont{A.~G.}~\bibnamefont{Lebed}},
  \bibinfo{book}{{\it Physics of Organic Superconductors and Conductors}},
  (\bibinfo{year}{Springer, Berlin, 2007}).



\bibitem[{\citenamefont{Pethick and Pines}(1986)}]{pet86}
\bibinfo{author}{\bibfnamefont{C.~J.} \bibnamefont{Pethick}} \bibnamefont{and}
  \bibinfo{author}{\bibfnamefont{D.}~\bibnamefont{Pines}},
  \bibinfo{journal}{Phys. Rev. Lett.} \textbf{\bibinfo{volume}{57}},
  \bibinfo{pages}{118} (\bibinfo{year}{1986}).



\bibitem[{\citenamefont{Vorontsov et~al.}(2005)}]{vorontsov2005}
\bibinfo{author}{\bibfnamefont{A.~B.}~\bibnamefont{Vorontsov}},
\bibinfo{author}{\bibfnamefont{J.~A.}~\bibnamefont{Sauls}}, \bibnamefont{and}
\bibinfo{author}{\bibfnamefont{M.~J.}~\bibnamefont{Graf}},
  \bibinfo{journal}{Phys. Rev. B} \textbf{\bibinfo{volume}{{\bf 72}}},
  \bibinfo{pages}{184501} (\bibinfo{year}{2005});
\bibinfo{author}{\bibfnamefont{A.~B.}~\bibnamefont{Vorontsov}} \bibnamefont{and}
\bibinfo{author}{\bibfnamefont{M.~J.}~\bibnamefont{Graf}},
  \bibinfo{journal}{ibid.} \textbf{\bibinfo{volume}{{\bf 74}}},
  \bibinfo{pages}{172504} (\bibinfo{year}{2006});
\bibinfo{author}{\bibfnamefont{J.}~\bibnamefont{Alexander}},
  \bibinfo{author}{\bibfnamefont{T.}~\bibnamefont{Orlando}},
  \bibinfo{author}{\bibfnamefont{D.}~\bibnamefont{Rainer}}, \bibnamefont{and}
  \bibinfo{author}{\bibfnamefont{P.}~\bibnamefont{Tedrow}},
  \bibinfo{journal}{ibid.} \textbf{\bibinfo{volume}{31}},
  \bibinfo{pages}{5811} (\bibinfo{year}{1985}).



\bibitem[{\citenamefont{Klein}(2004)}]{kle04}
\bibinfo{author}{\bibfnamefont{U.}~\bibnamefont{Klein}},
  \bibinfo{journal}{Phys. Rev. B} \textbf{\bibinfo{volume}{69}},
  \bibinfo{pages}{134518} (\bibinfo{year}{2004}).



\bibitem[{\citenamefont{Maki and Won}(1996)}]{mak96}
\bibinfo{author}{\bibfnamefont{K.}~\bibnamefont{Maki}} \bibnamefont{and}
  \bibinfo{author}{\bibfnamefont{H.}~\bibnamefont{Won}},
  \bibinfo{journal}{Czech. J. Phys.} \textbf{\bibinfo{volume}{46}},
  \bibinfo{pages}{1035} (\bibinfo{year}{1996}).



\bibitem[{\citenamefont{Shimahara and Rainer}(1997)}]{shi97}
\bibinfo{author}{\bibfnamefont{H.}~\bibnamefont{Shimahara}} \bibnamefont{and}
  \bibinfo{author}{\bibfnamefont{D.}~\bibnamefont{Rainer}},
  \bibinfo{journal}{J. Phys. Soc. Japan} \textbf{\bibinfo{volume}{66}},
  \bibinfo{pages}{3591} (\bibinfo{year}{1997});
\bibinfo{author}{\bibfnamefont{H.}~\bibnamefont{Shimahara}},
  \bibinfo{journal}{ibid.} \textbf{\bibinfo{volume}{67}},
  \bibinfo{pages}{736} (\bibinfo{year}{1998}).



\bibitem[{\citenamefont{Yang and Sondhi}(1998)}]{yan98}
\bibinfo{author}{\bibfnamefont{K.}~\bibnamefont{Yang}} \bibnamefont{and}
  \bibinfo{author}{\bibfnamefont{S.}~\bibnamefont{Sondhi}},
 \bibinfo{journal}{Phys. Rev. B} \textbf{\bibinfo{volume}{57}},
  \bibinfo{pages}{8566} (\bibinfo{year}{1998}).



\bibitem[{\citenamefont{Vorontsov}(2008)\citenamefont{Vorontsov
  unpublished}}]{abv08}
\bibinfo{author}{\bibfnamefont{A.~B.}~\bibnamefont{Vorontsov}},
  \bibnamefont{et al.},
  \bibinfo{journal}{(unpublished)}.



\bibitem[{\citenamefont{Burkhardt and Rainer}(1994)}]{bur94}
\bibinfo{author}{\bibfnamefont{H.}~\bibnamefont{Burkhardt}} \bibnamefont{and}
  \bibinfo{author}{\bibfnamefont{D.}~\bibnamefont{Rainer}},
  \bibinfo{journal}{Ann. Phys.} \textbf{\bibinfo{volume}{3}},
  \bibinfo{pages}{181} (\bibinfo{year}{1994}).



\bibitem[{\citenamefont{Wang et~al.}(2007)\citenamefont{Wang, Hu, and
  Ting}}]{wan07}
\bibinfo{author}{\bibfnamefont{Q.}~\bibnamefont{Wang}},
  \bibinfo{author}{\bibfnamefont{C.-R.} \bibnamefont{Hu}}, \bibnamefont{and}
  \bibinfo{author}{\bibfnamefont{C.-S.} \bibnamefont{Ting}},
  \bibinfo{journal}{Phys. Rev. B} \textbf{\bibinfo{volume}{75}},
  \bibinfo{pages}{184515} (\bibinfo{year}{2007}).

\bibitem{doped115}
\bibinfo{author}{\bibfnamefont{L.D.}~\bibnamefont{Pham}, et.al.},
  \bibinfo{journal}{Phys. Rev. Lett.} \textbf{\bibinfo{volume}{97}},
  \bibinfo{pages}{056404} (\bibinfo{year}{2006});
\bibinfo{author}{\bibfnamefont{E.D.}~\bibnamefont{Bauer}, et.al.},
  \bibinfo{journal}{Phys. Rev. B} \textbf{\bibinfo{volume}{73}},
  \bibinfo{pages}{245109} (\bibinfo{year}{2006});
\bibinfo{author}{\bibfnamefont{Y.}~\bibnamefont{Tokiwa}, et.al.}, 
  \bibinfo{journal}{arxiv:0804:2454} (\bibinfo{year}{unpublished}).



\end{thebibliography}
\end{document}